\documentclass{llncs}

\usepackage{amssymb}
\usepackage{amsmath}
\usepackage{graphicx}
\usepackage{color}

\newcommand{\mket}[1]{| #1 \rangle}
\newcommand{\mbra}[1]{\langle #1 |}

\newcommand{\keywords}[1]{\par\addvspace\baselineskip\noindent\keywordname\enspace\ignorespaces#1}

\begin{document}


\mainmatter

\title{GPGPU based simulations for one and two dimensional quantum walks}

\author{Marek Sawerwain \and Roman Gielerak}
\authorrunning{Marek Sawerwain \and Roman Gielerak}
\tocauthor{Marek Sawerwain and Roman Gielerak}

\institute{Institute of Control \& Computation Engineering \\
University of Zielona G\'ora, ul. Podg\'orna 50, Zielona G\'ora 65-246, Poland \\
\email{{M.Sawerwain,R.Gielerak}@issi.uz.zgora.pl}
}

\maketitle

\begin{abstract}
Simulations of standard 1D and 2D quantum walks have been performed within Quantum Computer Simulator (QCS system) environment and with the use of GPU supported by CUDA technology. In particular, simulations of quantum walks may be seen as an appropriate benchmarks for testing calculational power of the processors used. It was demonstrated by a series of tests that the use of CUDA based technology radically increases the computational power compared to the standard CPU based computations.
\end{abstract}

\section{Introduction}

Recently the concept of quantum walks attracted a big attention as they provide us with a very promising source of ideas for constructing new quantum algorithms \cite{Szegedy}. In particular exponential speedups of some classical problems have been discovered again (like in the Shor's algorithm case), the exponentially faster hitting described in \cite{Childs2} and \cite{Kempe2003b} seems to be a good example for this. Additionally, certain although not so particular as exponential speedups are, speedups of some classical problems like k-distinction problem, triangle and k-clique algorithms are examples of them \cite{Ambainis},\cite{Szegedy},\cite{Wocjan} have been obtained by the use of quantum walks concept. Another inspirations for studying quantum walks based applications may come from the observation \cite{Childs} that quantum walks might play the role of universal quantum calculations tool.

It is the main purpose of the present contribution to test some particular properties of quantum walks by using certain simulating tools. The main tools used for our simulations are the Zielona G\'ora Quantum Computer Simulator (as the main tool) and  the GPGPU equipped with new computational technology offered by CUDA (Compute Unified Device Architecture).

It is one of the main result of this note the demonstration of powerful incrementation of calculational abilities if we use GPGPU of new generation compared with standard CPU based computations.

Organisation of this note is as follows: in Section~(\ref{lbl:some:mathematical:preliminaries}) we provide a reader with the basic definitions and constructions connected with quantum walks on general graphs. Numerical implementations of quantum walks on certain lattice structures coming from that of $\mathbb{Z}^1$ and $\mathbb{Z}^2$ together with the corresponding numerical algorithms are presented in Section~(\ref{lbl:algorithm:for:simulating:qwalks:GPGPU}). Additionally several examples of our simulations will be presented and discussed briefly there.

\section{Some mathematical preliminaries}\label{lbl:some:mathematical:preliminaries}

Let $G=(\mathbb{V}, \mathbb{E}, l)$ be a graph and where as obviously $\mathbb{V}$ stands for the set of vertices of $G$ the number of elements $|\mathbb{V}|$ of which is equal to $N$, $\mathbb{E}$ stands for the set of (undirected) edges of $G$ and $l : \mathbb{E} \rightarrow \{\mathbb{V}, \mathbb{V}\}$ is the edges labelling function. The corresponding incidence matrix of $G$ will denoted as $M_G$. With the use of $M_G$ the continuous time Markov walk on $G$ can be described by the corresponding, canonical Markovian transition semi-group $T_{t}=\exp(-t M_G)$ and (eventually) an initial distribution $\pi_{0}$. For any $v \in \mathbb{V}$ we denote by $d(v)$ the corresponding degree of vertex $v$ and let then $\mathbb{P}(v)=(P_1,P_2,\ldots,P_{d(v)})$ be discrete probability measure assigning a probability that the walker jump with probability $p_i$ by the use of i-th edge $e_i$ connecting the vertex $v$ with $l(e_i)(2)=\omega$. The system $(\mathbb{P}, \pi_0)$, where $\pi_0$ is an initial distribution gives rise to the discrete step random walk on $G$.

The corresponding quantum walks on $G$ can be constructed in the following way. 

The continuous time quantum walk on $G$: by the very definition starting from the vertex $v_0$ (with probability $\pi_0$) after time $t$ we arrive at the vertex obtained by the action of the unitary group $U^G_t=\exp(i t H_G)$ (where $H_G$ stands for the corresponding graph Hamiltonian) acting in the Hilbert space $\mathcal{H}^G = \oplus_{v \in \mathbb{V}} \mket{v}$. However only the discrete time processes will be discussed in this note.

\subsection{The discrete time Markovian quantum walks}

Let$\mathcal{H}^G = \oplus_{v \in \mathbb{V}} \mket{v}$ be the canonical Hilbert space associated with $G$, obviously $\mathcal{H}^G \simeq \mathbb{C}^{N}$. For any $v \in \mathbb{V}$ let $d(v)$ be degree of $v$. Then the local Hilbert space $\mathcal{H}_{v}$ is defined as a space isomorphic to $\mathbb{C}^{d(v)}$, explicitly $\mathcal{H}_{v} = \oplus_{e_u}\mket{e_u}$, $e_u$ runs over all edges connecting the vertex $v$ with others. A collections $\mathbb{C}=(C_v, v \in \mathbb{V})$ of unitary maps acting on the spaces $\mathcal{H}_{v}$, $v \in \mathbb{V}$ and fulfilling additionally certain natural coincidence conditions, see i.e. \cite{Kempe2003b}; will be called a "coin flip transformation sequence". In other words, for any $v \in \mathbb{V}$:
\begin{equation}
\begin{array}{cl}
C_v :				& \mket{v} \otimes \mathcal{H}_{v} \rightarrow \mket{v} \otimes \mathcal{H}_{v} \\ 
\mathrm{where}	& \mket{v} \otimes \mket{\omega} \rightarrow \mket{v} \otimes C_v \mket{\omega}
\end{array}
\end{equation}

The global Hilbert space $\mathcal{H}$ is defined as $\mathcal{H}=\oplus_{v \in \mathbb{V}} \mket{v} \otimes \mathcal{H}_{v}$ and the corresponding discrete quantum walk on $G$, providing the family $\mathbb{C}\;$ is given, can be defined by its one step transformation:
\begin{equation}
U = S ( I \otimes \mathbb{C} \; )
\end{equation}
where the shift transformation $S$ is defined as:
\begin{equation}
S\mket{v,e}=\mket{v',e} \;\;\; \mathrm{if} \;\;\; l(e)=\{v,v'\} .
\end{equation}

Several questions (with analogy to the classical case, especially the problems connected to the mixing and hitting times on a large class of graphs have been studied intensively) can be studied, the question about limiting probability distributions and hitting times are the most popular one \cite{Kempe2003a}. 

Although intensive simulations of quantum walks on many complex graphs are planned to be done we have concentrated first on some simplest quantum walks on the infinite (the finite amount of with appropriate boundary condition are considered in real simulation tasks of course) graphs $\mathbb{Z}^1$ and $\mathbb{Z}^2$ that we describe now.

\subsection{Quantum walks on lattice $\mathbb{Z}^1$ and on lattice $\mathbb{Z}^2$}

With the lattice $\mathbb{Z}^1$ we associate the Hilbert space $l_2(\mathbb{Z})=\oplus_{n \in \mathbb{Z}}\mket{n}$ and the coin flip transformation $\mathbb{C}$ acting in $\mathbb{C}^{2} \equiv \mket{R} \oplus \mket{L}$ symbolising the possible steps in the right $(\mket{R})$ or left direction $(\mket{L})$ is given. Then the corresponding Hilbert space $l_2(\mathbb{Z})  \otimes \mathbb{C}^2$ can be seen as a space of infinite sequences $\mket{\psi} \approx ((\alpha_j,k), \alpha_j \in \mathbb{Z}, \; k \in \{R,L\})$, i.e. any vector $\mket{\psi} \in l_2(\mathbb{Z}) \otimes \mathbb{C}^2$ can be given by:
\begin{equation}
\mket{\psi} = \sum_{j \in \mathbb{Z}, k=L,R} \alpha_{jk} \mket{j,k} \;\;\; \mathrm{where} \;\;\; \sum_{j,k} {|\alpha_{j,k}|}^2 = 1
\end{equation}
Different choices of $\mathbb{C}\;$ and shift operators lead to different models of quantum walks on the lattice $\mathbb{Z}^1$.

The graph Hilbert space $\mathcal{H}^{\mathbb{Z}^2}$ for the 2D lattice $\mathbb{Z}^2$ is defined as 
\begin{equation*}
\mathcal{H}^{\mathbb{H}_2} = \oplus_{j,m \in Z}\mket{j,m} = l_2(\mathbb{Z}^2) . 
\end{equation*}
The degrees of all vertices are equal to $4$ and therefore the local Hilbert spaces are isomorphic with $\mathbb{C}^4$ to be identified with R,L,U,D (right, left, up, down) steps on the lattice. The total space $\mathcal{H}=l_2(\mathbb{Z}^2) \otimes \mathbb{C}^4$ and the typical vector $\mket{\psi} \in \mathcal{H}$ can be decomposed as 

\begin{equation}
\mket{ \psi } = \sum_{j,k=1}^{4} \sum_{m,n = -\infty}^{+\infty} \alpha_{j,k,m,n}\mket{j,k}\mket{m,n}
\;\;\; \mathrm{with} \;\;\;
\sum_{j,k=1}^{4} \sum_{m,n = -\infty}^{+\infty} {|\alpha_{j,k,m,n}|}^2 = 1
\end{equation}

A different versions of the corresponding coin flip transformation $\mathbb{C}$ and the shift operators (reflecting some additional topological constraints) then lead to different quantum walk models on $\mathbb{Z}^2$ lattice. Some of them will be presented for simulations performed in the next section including some quantum walk models on $\mathbb{Z}^1$ as well.

\section{The algorithm for simulating quantum walks on GPGPU}\label{lbl:algorithm:for:simulating:qwalks:GPGPU}

The calculation routine for simulation of quantum walks can be build directly from the definition of state of quantum walker walking on the lattice $\mathbb{Z}^2$. In general a state of the quantum walker at time $t$ is given in the following way:

\begin{equation}
\mket{ \psi(t) } = \sum_{j,k=0}^{1} \sum_{m,n = -\infty}^{+\infty} \alpha_{j,k,m,n}(t)\mket{j,k}\mket{m,n}
\label{lbl:eq:quantum:walker:state}
\end{equation}
where $\alpha_{j,k,m,n}(t) \in \mathbb{C} \;\; \mathrm{and} \; \sum_{j,k}\sum_{m,n} {|\alpha_{j,k,m,n}(t)|}^2 = 1$. The evolution of the quantum walker system over time $t$ is expressed by following unitary operator
\begin{equation}
U = S (C \otimes I)
\end{equation}
where $S$ is the shift operator, $I$ represents the identity operator and $C$ is the coin operator (in most cases we can  assume that the coin is represented by Hadamard operator, but there exist other representations of the coin operator e.g. Fourier and Grover coins) which acts on the local $\mathcal{H}_2 \otimes \mathcal{H}_2$ subspace of whole walker Hilbert space system $l_2(\mathbb{Z}^2) \otimes \mathcal{H}_2 \otimes \mathcal{H}_2$. 

In this contribution we propose rather special definition of shift operator for two-dimensional quantum walks. A comparison of our definition with those used frequently can be found in \cite{Oliviera} and \cite{QWalk2008}. We also use the random broken links (termed RBL) technique, first developed for one dimensional quantum walks and introduced in \cite{Romanelli}. The RBL technique was generalised for two-dimensional case in \cite{Oliviera}.

The used definition of shift operator which coincides with physical and mathematical lattice is the following  
\begin{align}
S = \sum_{j,k=0}^{1} \sum_{m,n=-\infty}^{+\infty} \mket{j,k}\mbra{j,k} \otimes \mket{m+(-1)^j(1-\delta_{j,k}),n+(-1)^j\delta_{j,k}}\mbra{m,n}
\label{lbl:eq:shift:operator}
\end{align}
where $\delta_{j,k}$ is Dirac discrete delta function. 

\begin{figure}[!ht]
\includegraphics[width=6.5cm]{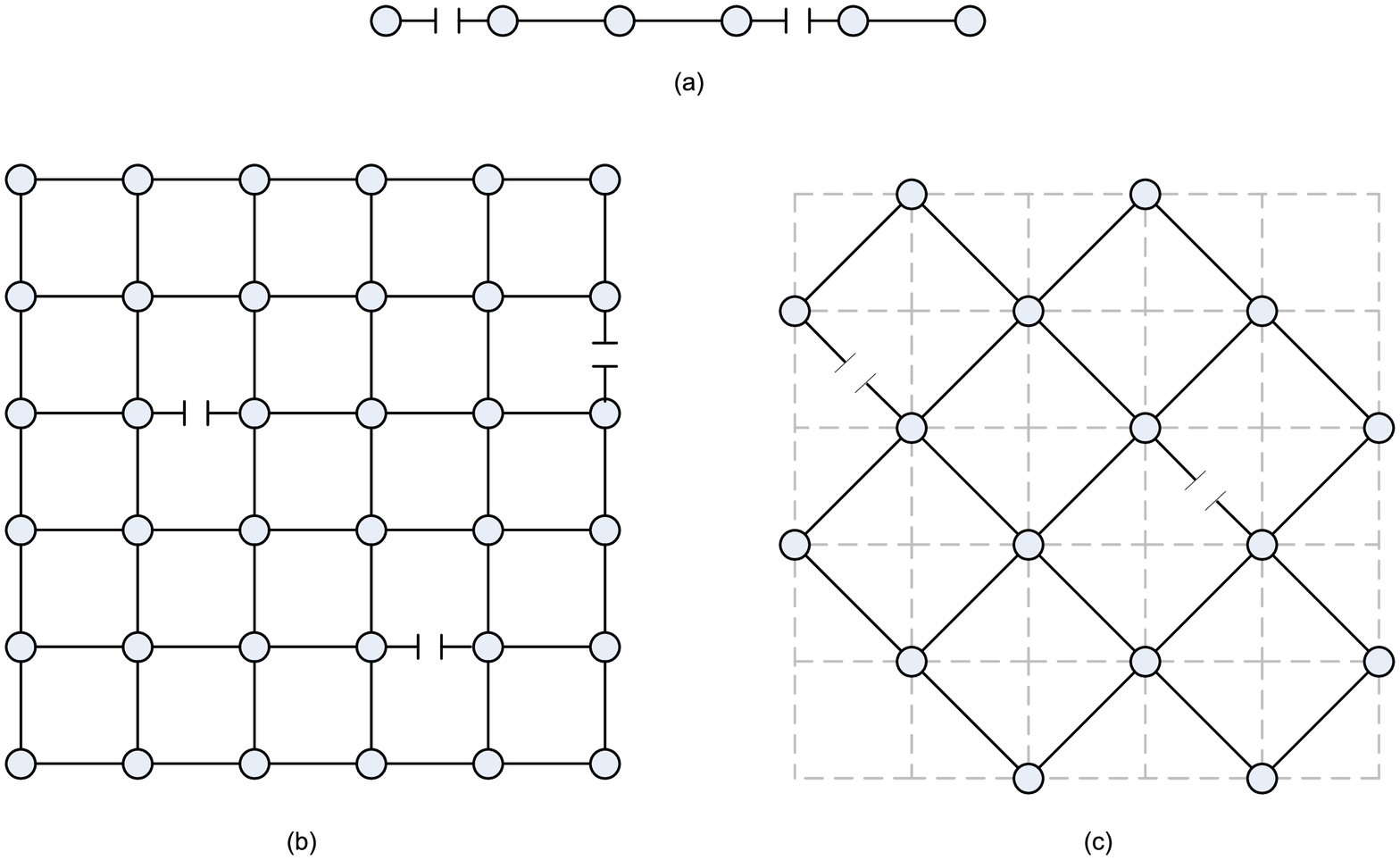}
\centering
\caption{The idea of broken links for one-dimensional quantum walk (a) where we show a broken links between sites, part (b) and part (c) represent examples of natural and diagonal lattices for two-dimensional quantum walk}
\label{lbl:fig:lattices}
\end{figure}

The following function includes the possibility of appearance of the broken line in a path in between site $(j,k)$ and $(m,n)$ in example depicted on Fig.~\ref{lbl:fig:lattices}:
\begin{equation}
\mathcal{RBL} (j, k, m, n) = \left\{ \begin{array}{lcl}
{(-1)}^j	& -	& \mathrm{if \; link \; to} \; m+(-1)^j(1-\delta_{j,k}), \\
			&		& \;\;\; n+(-1)^j \delta_{j,k} \; \mathrm{is \; closed} \\
0			& -	& \mathrm{if \; link \; is \; open}
\end{array} \right.
\end{equation}
where $j,k \in \{ 0, 1 \}$.

After applying shift operator (\ref{lbl:eq:shift:operator}) to state (\ref{lbl:eq:quantum:walker:state}) the evolution can be summarised in following way:
\begin{align}
\psi_{(1-j,1-k,m,n)}(t+1) = \sum_{j',k'=0}^{1} C_{j+\mathcal{RBL} (j,k,m,n),k \oplus \mathcal{RBL} (j,k,m,n),j',k'} \notag \\
\cdot \; \psi_{(j',k',m+\mathcal{RBL}(j,k,m,n)(1-\delta_{j,k}),n + \mathcal{RBL}(j,k,m,n))\delta_{j,k}}(t)
\label{lbl:eqn:evolution:equation}
\end{align}
where $\oplus$ represents addition modulo two.

The algorithm to simulate two-dimensional quantum walks is directly basing on the evolution given by equation (\ref{lbl:eqn:evolution:equation}). It can be implemented on the traditional architecture for standard CPU and of course for GPGPU based solutions.

Figure (\ref{lbl:fig:code:1d:segment}) shows a fragment of calculation routine for GPGPU which task is to compute values for the next iteration of quantum walk in the segment case. In each iteration all points in the segment attain a new values. This means that the all available GPGPU cores are fully used. Additionally, the efficiency can be increased by better usage of threads available in CUDA architecture.

\begin{figure}
{\small
\begin{verbatim}
__global__ void one_iteration_segment( 
            cuFloatComplex *A0, cuFloatComplex *A1,
            cuFloatComplex *Atemp0, cuFloatComplex *Atemp1, int *RBL0,
            int *RBL1, cuFloatComplex *C, int N) {
    int m = blockIdx.x * blockDim.x + threadIdx.x;
    int L, cidx1, cidx2;

    if (m<N) {
        L = RBL0[m]; cidx1=(L*2); cidx2=(L*2)+1;
        Atemp1[m] = cuCaddf(cuCaddf(Atemp1[m], cuCmulf(C[cidx1], A0[m+L])),
                            cuCaddf(Atemp1[m], cuCmulf(C[cidx2], A1[m+L])));

        L = RBL1[m]; cidx1=((1+L)*2); cidx2=((1+L)*2)+1;
        Atemp0[m] = cuCaddf(cuCaddf(Atemp0[m], cuCmulf(C[cidx1], A0[m+L])),
                            cuCaddf(Atemp0[m], cuCmulf(C[cidx2], A1[m+L])));
    }
}
\end{verbatim}
\centering
}
\caption{The kernel function to compute trajectories of one-dimensional quantum walks on segment}
\label{lbl:fig:code:1d:segment}
\end{figure}

\begin{figure}
\begin{tabular}{cc}
\includegraphics[width=6.0cm]{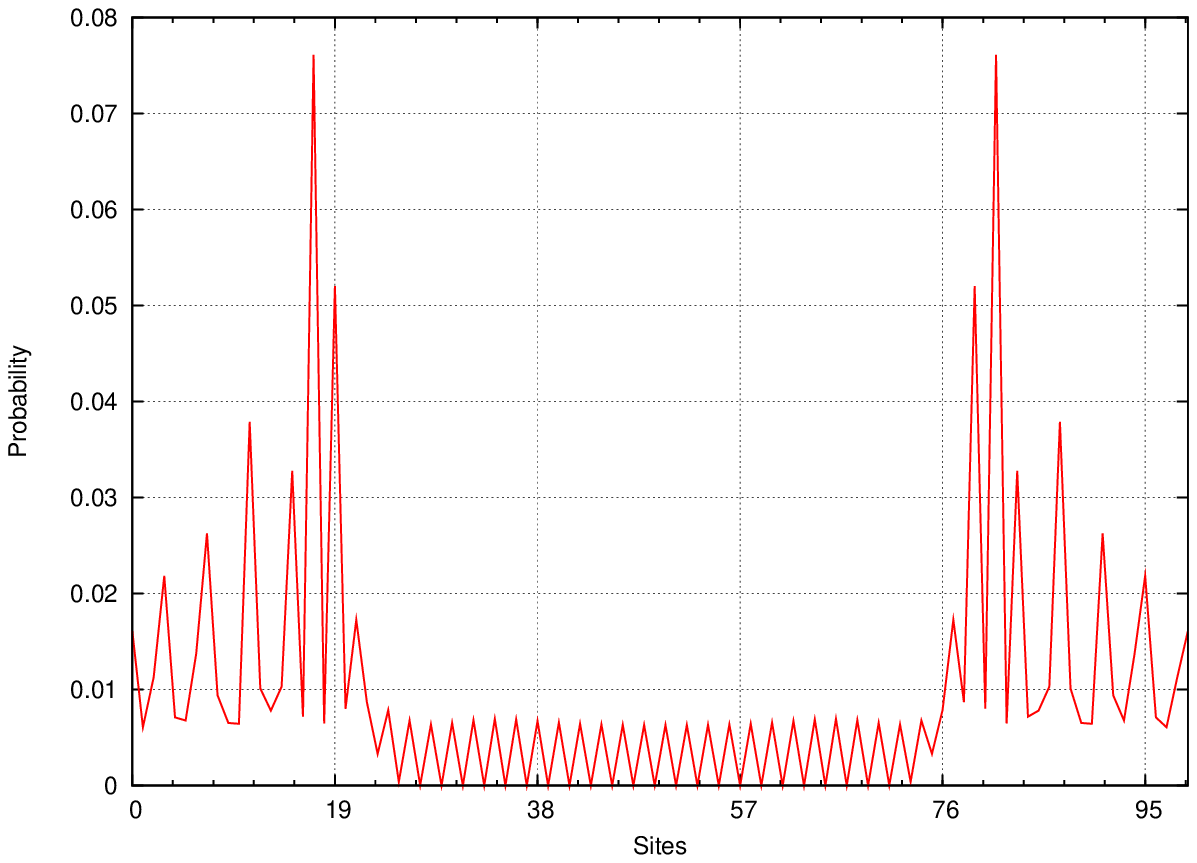} & \includegraphics[width=6.0cm]{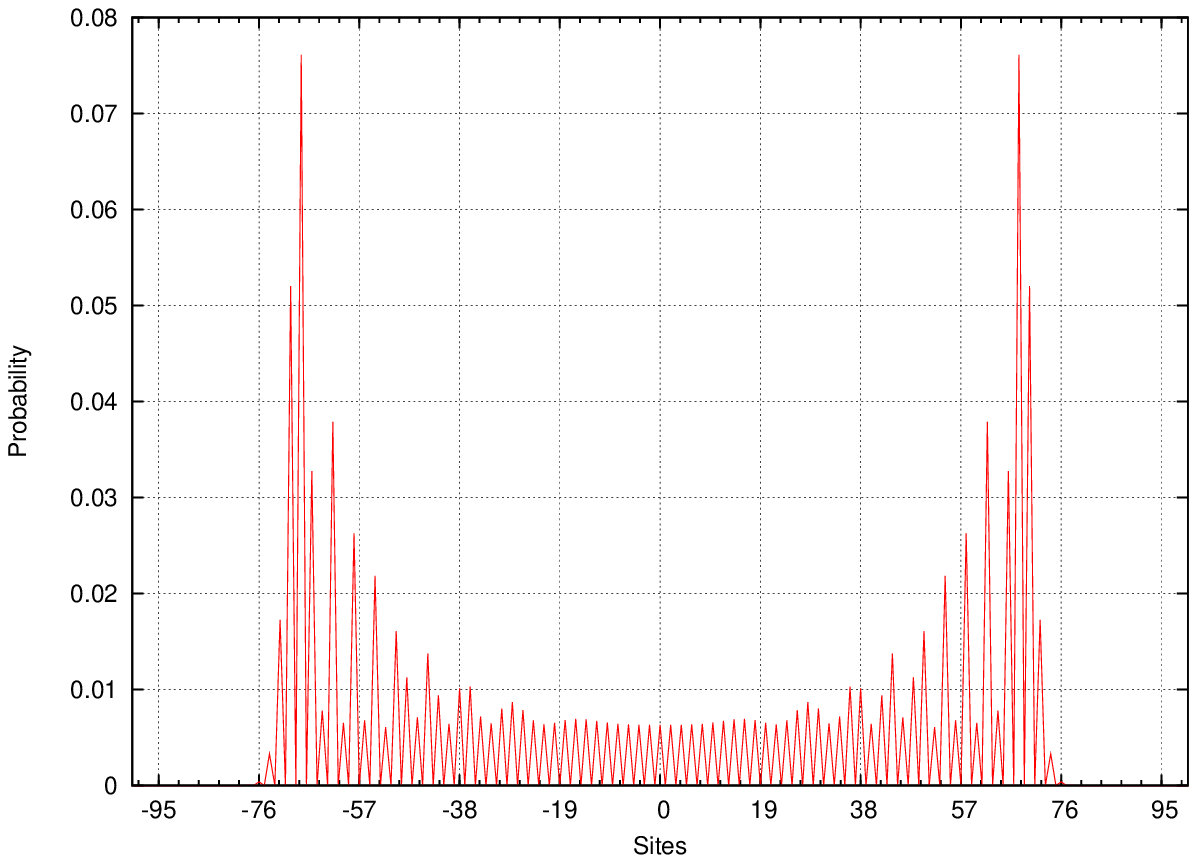} \\
\end{tabular}
\centering
\caption{The one-dimensional quantum walk on segment (left plot) and on line (right plot). In both cases amount of iterations are the same and equal to 100}
\label{lbl:fig:1d:qwalk}
\end{figure}

The function to calculate the probability distribution for one-dimensional quantum walk on the line is very similar to the segment case with one important difference. In the $i$-th iteration the quantum walker cannot be farther than $i$ sites from its initial position. The necessary change in GPGPU routine is expressed as
{\small
\begin{verbatim}
int m = blockIdx.x * blockDim.x + threadIdx.x;
int left, right;
left = max(midpoint - extra - iteration, 1); 
right = min(midpoint + extra + iteration, N-1);
if (m>=left && m<=right) { ... }
\end{verbatim}
}
The probability distributions connected to a calculated trajectory for segment and line are not the same, which is illustrated on the Fig.~(\ref{lbl:fig:1d:qwalk}).

\begin{figure}
\begin{tabular}{cc}
(a) & (b) \\
\includegraphics[width=5.5cm]{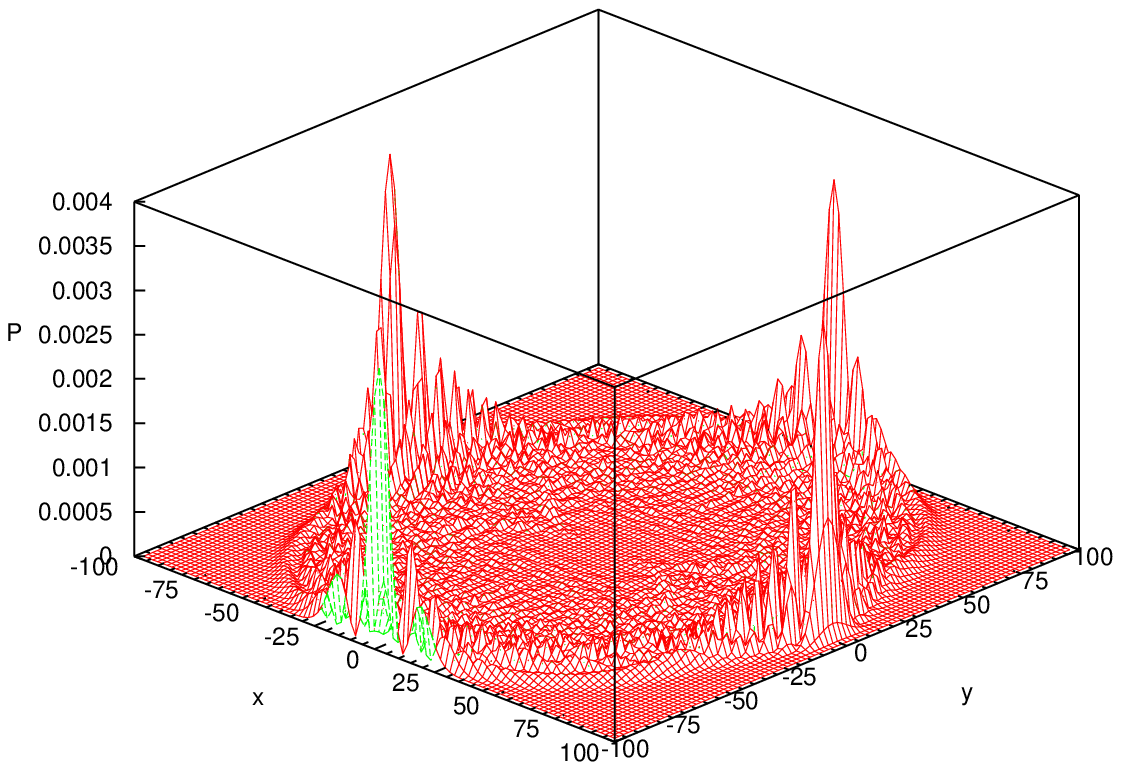} & \includegraphics[width=5.5cm]{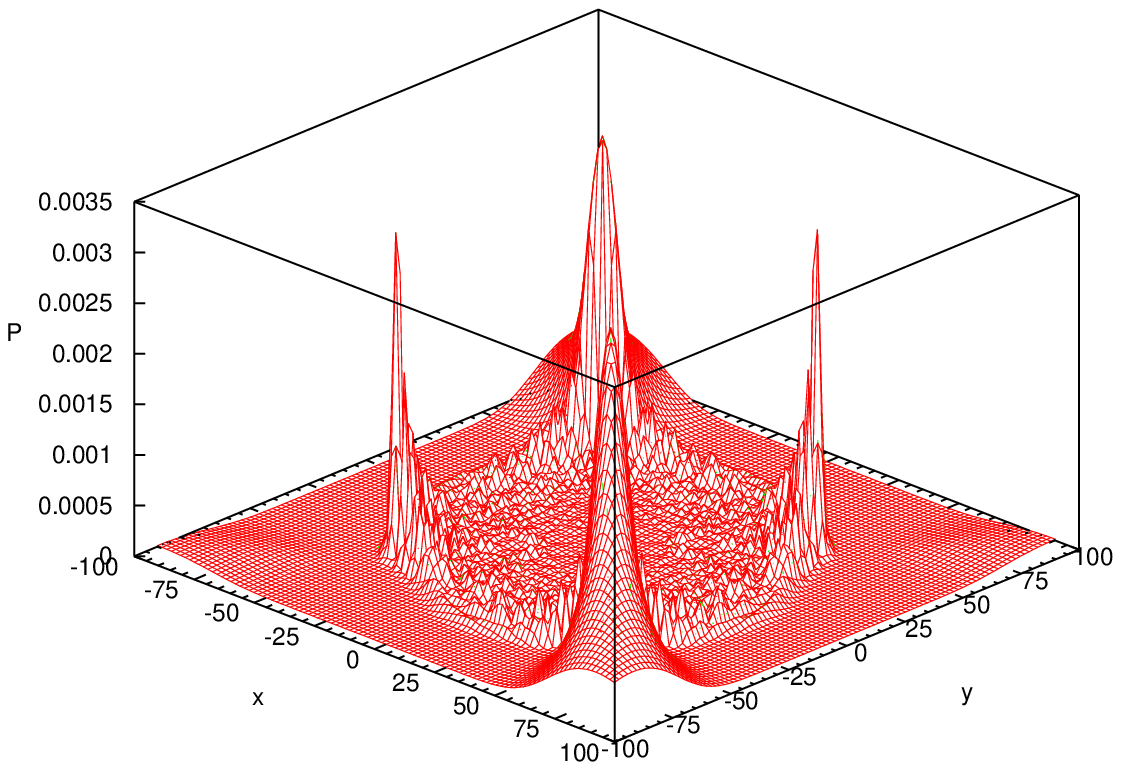}
\end{tabular}
\centering
\caption{Probability distribution of two-dimensional quantum walks with Fourier walker and Fourier's initial state on the diagonal lattice depicted on plot (a) and natural lattice depicted on plot (b)}
\end{figure}

\subsection{Complexity analysis briefly}

The computational complexity $T^{1d}$ of simulation of quantum walk for one-dimensional case strictly depends on the size of segment or line used. Let $N$ be a size of segment for arising in $i$-th iteration, then:
\begin{equation}
T^{1d}_{i}(N) = \sum_{m=0}^{N-1} \sum_{j=0}^{1} \left( T_{RBL_m} \sum_{k=0}^{1} \left( T_{OP_m} \right) \right) = 4 \cdot T_{RBL_m} \cdot T_{OP_m} \cdot N
\label{lbl:eq:complexity:segment}
\end{equation}
It is possible to write equation (\ref{lbl:eq:complexity:segment}) in such form because operations ${RBL_m}$ and ${OP_m}$ need the same constant time of work which is independent on the number of iteration. The symbol $T_{RBL_m}$ stands for amount of time necessary to process information about broken links, the second symbol $T_{OP_m}$ stands for amount of time necessary to process the probability amplitude of selected point. The use of GPGPU means, that the processing time can be divided by the number $N_c$ of available cores, because operations ${RBL_m}$ and ${OP_m}$ are independent for each point in the segment or line (as well as in two-dimensional case):
\begin{equation}
T^{1d}_{GPGPU}(n_i) = T^{1d}_{i}(N) / N_c
\end{equation}
The simulation of quantum walk on line shows one important difference comparing to segment case. The point is that the new values of points are calculated only in the partially and the size of simulated portion strictly depends on the number of iteration. The complexity can be denoted as (where $T_{RBL_m}$ and $T_{OP_m}$ have the same sense as in previous segment case):
\begin{equation}
	T^{1d}_{i}(N) = \sum_{m=\mathrm{l}_i}^{\mathrm{r}_i} \sum_{j=0}^{1} \left( T_{RBL_m}\sum_{k=0}^{1} \left( T_{OP_m} \right) \right) = 4 \cdot T_{RBL_m} \cdot T_{OP_m} \cdot ((\mathrm{r}_i-\mathrm{l}_i) + 1 )
\end{equation}
where the $l_i$ and $r_i$ are given by:
\begin{equation}
\mathrm{l}_i = \max(\mathrm{MP} - \mathrm{ES} - \mathrm{i}, 1), \;\;\; \mathrm{r}_i  =  \min(\mathrm{MP} + \mathrm{ES} + \mathrm{i}, N - 1) 
\end{equation}
where MP is the midpoint index of line, the value ES is used to pad and therefore prevents from range errors. However, these values are constant so the difference of $r_i$ and $l_i$ for $i$-th iteration can be expressed as
\begin{equation}
(r_i - l_i) = (2 \cdot i) + 1 .
\label{lbl:eq:difference}
\end{equation}
In the case of small systems (ten, twenty or fifty points), the equation (\ref{lbl:eq:difference}) shows that the most of available cores in GPGPU are not fully utilised. In the case of bigger systems this problem is not arising. The difference in (\ref{lbl:eq:difference}) for higher iteration number is bigger and what is more the values of this difference for all iterations form the arithmetic sequence, what means that in sense of complexity theory that only linear speedup is achieved, however for segment with size of 10000 points the obtained speedup is nearly hundredfold. The corresponding times have been depicted in Tables (\ref{lbl:tbl:1d:results}), (\ref{lbl:tbl:2d:results}) and (\ref{lbl:tbl:2d:mth:results:03}). 

\begin{table}
\caption{The measured times of calculations of one-dimensional quantum walks trajectories for segments with different sizes (without broken links) }
\begin{center}
\begin{tabular}{|c|c|c|c|}
\hline
		& Core 2 Duo 8400 (1 core)  & Geforce 9600 GT (64 cores)	 & Geforce 280 (240 cores) \\ \hline\hline
Size  & Time in ms 					 & Time in ms						 & Time in ms 					\\ \hline \hline
100	&	6.00							 & 4.51								 & 2.48							\\
1000	&	663.00						 & 40.186							 & 24.171						\\
5000	&	22685.00						 & 612.67							 & 281.57						\\
10000	&	96362.00						 & 2324.3002						 & 884.297						\\ \hline \hline
\end{tabular}
\end{center}\label{lbl:tbl:1d:results} 
\end{table}

\begin{table}
\caption{The measured times of calculations of two-dimensional quantum walks trajectories for diagonal lattice without broken links. The measured times for the case of Core 2 Duo and two-thread computational routine are presented in bracket}
\begin{center}
\begin{tabular}{|c|c|c|c|}
\hline
		& Core 2 Duo 8400 (1 core) & Geforce 9600 GT (64 cores)	& Geforce 280 (240 cores) \\ \hline\hline
Size  & Time in ms 					& Time in ms						& Time in ms \\ \hline \hline
100		&	700 (413)					& 195									& 64 						\\
200		&	5480 (2945)					& 1003								& 363						\\
300		&	19140	(10512)				& 3423								& 1137					\\ 
400		&	47230	(25785)				& 8123								& 3127					\\ 
500		&	92530	(50274)				& 14462								& 4706					\\ \hline \hline 
\end{tabular}
\end{center}
\label{lbl:tbl:2d:results} 
\end{table}

\begin{table}
\caption{The measured times of the calculations of two-dimensional quantum walks trajectories for diagonal lattice without broken links. The simulation was performed on two Intel Xeon E5420 2.50 Ghz processors, the multi-threaded calculation subroutine was compiled with GCC compiler with "-O3" option}
\begin{center}
\begin{tabular}{|c|c|c|c|c|}
\hline
			& (1-thread) 			& (2th)				& (4th)		& (8th)	\\ \hline\hline
Size 		& Time in ms 			& Time in ms			& Time in ms	&  Time in ms	\\ \hline \hline
100		&	770					& 	409					& 299				&	206			\\
200		&	6740					& 	3281					& 2190			&	1382			\\
300		&	22910					& 	12029					& 7130			&	4848			\\
400		&	56610					& 	29853					& 17699			&	12069			\\
500		&	114060				& 	65292					& 33275			&	24602			\\ \hline \hline 
\end{tabular}\label{lbl:tbl:2d:mth:results:03} 
\end{center}
\end{table}

The use of equation (\ref{lbl:eqn:evolution:equation}) allows to estimate computational complexity of two-dimensional quantum walk trajectory calculations in i-th iteration:
\begin{align}
	T^{2d}_{i}(N) = \sum_{m=\mathrm{lb}_{i}}^{\mathrm{rb}_{i}} \sum_{n=\mathrm{lb}_{i}}^{\mathrm{rb}_{i}} \sum_{j=0}^{1} \sum_{k=0}^{1} T_{RBL_{(m,n,j,k)}} 4 \cdot  T_{OP_{(m,n,j,k)}} = \notag \\
16 \cdot T_{RBL_{(m,n,j,k)}} \cdot T_{OP_{(m,n,j,k)}} \cdot ( (\mathrm{rb}_i - \mathrm{lb}_i) + 1)^2 ,
\label{lbl:eqn:2d:complexity}
\end{align}
where $N$ means the length of trajectory calculated.

The variables $\mathrm{lb}_{i}$ and $\mathrm{rb}_{i}$ have the same meaning as $l_i$,  $r_i$ introduced before and are calculated in the following way:
\begin{equation}
\mathrm{\mathrm{lb}_{i}} = \max(\mathrm{MP} - \mathrm{ES} - \mathrm{i}, 1), \;\;\; \mathrm{\mathrm{rb}_{i}} = \min(\mathrm{MP} + \mathrm{ES} + \mathrm{i}, 2 \cdot \mathrm{MP} - 1) .
\end{equation}

\begin{figure}
\begin{tabular}{c}
\includegraphics[height=4.0cm]{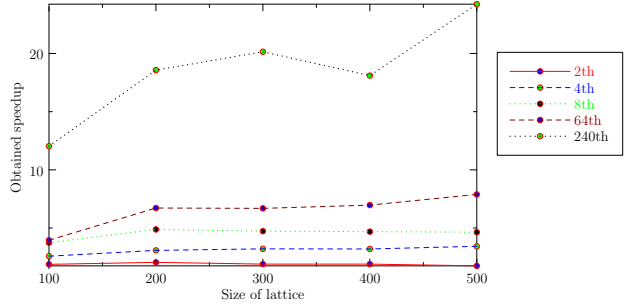} \\
\end{tabular}
\centering
\caption{The obtained values of speedup for simulations of two dimensional quantum walks on diagonal lattice. Graphs labelled as 2th, 4th and 8th are compared to one-thread computational routine. Graphs marked as 64th and 240th represent the speedup obtained by computational routine executed on the Geforce video card 9600 GT and GTX 280 respectively}
\label{lbl:fig:obtained:speedup}
\end{figure}

The expression $( (\mathrm{rb}_i - \mathrm{lb}_i) + 1)^2$ in equation (\ref{lbl:eqn:2d:complexity}) can be expressed in the following way
\begin{equation}
( (\mathrm{rb}_i - \mathrm{lb}_i) + 1)^2 = ((2 \cdot i ) + 1)^2 .
\end{equation}
Speedups obtained for 2D quantum walks are presented in Fig.~(\ref{lbl:fig:obtained:speedup}) are based on results presented in Tables~(\ref{lbl:tbl:2d:results}) and (\ref{lbl:tbl:2d:mth:results:03}).

\section{Conclusions and further work}

The specialised software to simulate one and two dimensional random quantum walks without and with broken links  has been presented in this article. The used software is a part of the Quantum Computing Simulator presented in \cite{MSawerwain2008}.
A significant speedup of the simulations process comparing to previous paper \cite{QWalk2008} have been achieved. The used technologies enables to simulate effectively much more complex quantum walks then previously known. Additionally, certain more deeper notions connected to the analysis of quantum  walks behaviour can be analysed using computer  simulations as an appropriate tool.

\subsubsection*{Acknowledgments}

We acknowledge useful discussions on the QCS with the \textit{Q-INFO} group at the Institute of Control and Computation Engineering of the University of Zielona G\'ora, Poland.


\begin{thebibliography}{99}
\bibitem{Aharonov2001} Aharonov~D., Ambainis~A., Kempe~J. and Vazirani~U., Quantum walks on graphs, in {\em Proceedings of 33th STOC}, pages 50--59, ACM, 2001.

\bibitem{Ambainis} Ambainis~A.: {Quantum walks and their algorithmic applications}, International Journal of Quantum Information, Vol.~1, No.~4, pp.:~507--518, 2003.

\bibitem{Childs} Childs~A.M.: {Universal computation by quantum walk}, Phys. Rev. Lett., Vol.~102, pp.~180501, 2009.

\bibitem{Childs2} Childs~A.M., Cleve R., Deotto E., Farhi E., Gutmann S., Spielman D.A., {Exponential algorithmic speedup by quantum walk}, 	Proc. 35th ACM Symposium on Theory of Computing (STOC 2003), pp. 59-68.

\bibitem{Kempe2003a} Kempe~J.: {Quantum random walk algorithms}, Contemp. Phys. 44 (3), pp.~302–327, 2003.

\bibitem{Kempe2003b} Kempe~J.: {Quantum random walks hit exponentially faster}, in {\em Proceedings of 7th Intern. Workshop on Randomization and  Approximation Techniques in Computer Science}, LNCS, Springer, Heidelberg, pp.~354--369, 2003, see also at arXiv:quant-ph/0205083. 

\bibitem{Kendon2006} Kendon~V., {Decoherence in quantum walks: a review}, Math. Struct. in Comp. Sci 17(6), pp.~1169-1220, 2006, arXiv:quant-ph/0606016v3. 

\bibitem{QWalk2008} Marquezino~F.L., Portugal~R., {The QWalk Simulator of Quantum Walks}, Computer Physics Communications, 
Vol.~179, Issue~5, pp.~359--369, 2008, see also arXiv:quant-ph/0803.3459v1.

\bibitem{Nayak} Nayak~A., Vishwanath~A.: {Quantum Walk on the Line}, arXiv:quant-ph/0010117.

\bibitem{Oliviera} Oliveira~A.C., Portugal~R., Donangelo~R., Decoherence in two-dimensional quantum walks, Phys. Rev. A 74, 012312.

\bibitem{AmandaSlit2007}
Oliveira~A., Portugal~R., and Donangelo~R., Simulation of the single- and double-slit experiments with quantum walkers, arXiv:quant-ph/0706.3181, 2007.

\bibitem{Romanelli} Romanelli~A., Siri~R., Abal~G., Auyuanet~A., and Donangelo~R., Physica A 347C (2005), arXiv:quant-ph/0403192.
 
\bibitem{MSawerwain2008} Sawerwain M.: {Parallel algorithm for simulation of circuit and one-way quantum computation models}, In: R.Wyrzykowski et al., eds, Parallel Processing and Applied Mathematics, Proc. 7th Int.
Conf., PPAM 2008, Gda\'nsk, Poland, Vol.~4967, pp.~530--539.

\bibitem{MSawerwainRGielerak2008} Sawerwain~M., Gielerak~R., Natural quantum operational semantics with predicates, Int. J. Appl. Math. Comput. Sci., 2008, Vol. 18, No. 3, pp.~341-–359.

\bibitem{Szegedy} Szegedy~M.: {Quantum Speedup of Markov Chain Based Algorithms}, Proc. of 45th Annual IEEE Symposium on Foundations of Computer Science, pp.:~32--41, 2004.

\bibitem{Wocjan} Wocjan~P., Abeyesinghe~A., {Speedup via quantum sampling}, Phys. Rev. A 78, 042336, 2008.

\end{thebibliography}
\end{document}